\documentclass[sigconf,screen]{acmart}
\usepackage{graphicx}
\usepackage{xr}
\usepackage{catchfilebetweentags}
\usepackage{multirow}
\usepackage{booktabs}
\usepackage{xspace}
\usepackage{enumitem}
\usepackage[caption=false]{subfig}
\usepackage{balance}
\usepackage{microtype}

\graphicspath{{figs/}}

\newcommand{\ours}{\textsc{RefBERT}\xspace}
\newcommand{\oursCE}{\textsc{RefBERT} w/o BoT \& CL\xspace}
\newcommand{\oursCECL}{\textsc{RefBERT} w/o BoT\xspace}
\newcommand{\oursCEBOT}{\textsc{RefBERT} w/o CL\xspace}
\newcommand{\oursBOTCL}{\textsc{RefBERT} w/o cMLM\xspace}

\setcopyright{acmlicensed}
\acmPrice{15.00}
\acmDOI{10.1145/3597926.3598092}
\acmYear{2023}
\copyrightyear{2023}
\acmSubmissionID{issta23main-p208-p}
\acmISBN{979-8-4007-0221-1/23/07}
\acmConference[ISSTA '23]{Proceedings of the 32nd ACM SIGSOFT International Symposium on Software Testing and Analysis}{July 17--21, 2023}{Seattle, WA, United States}
\acmBooktitle{Proceedings of the 32nd ACM SIGSOFT International Symposium on Software Testing and Analysis (ISSTA '23), July 17--21, 2023, Seattle, WA, United States}
\received{2023-02-16}
\received[accepted]{2023-05-03}

\begin{document}

\title{\ours: A Two-Stage Pre-trained Framework for\\Automatic Rename Refactoring}

\author{Hao Liu}
\affiliation{%
  \department{Key Laboratory of Multimedia Trusted Perception and Efficient Computing, Ministry of Education of China}
  \country{School of Informatics, Xiamen University}
}
\email{haoliu@stu.xmu.edu.cn}

\author{Yanlin Wang}
\affiliation{%
  \institution{School of Software Engineering\\Sun Yat-sen University}
  \country{China}
}
\email{wangylin36@mail.sysu.edu.cn}

\author{Zhao Wei}
\author{Yong Xu}
\author{Juhong Wang}
\affiliation{%
  \country{Tencent}
}
\email{{zachwei, rogerxu, julietwang}@tencent.com}

\author{Hui Li}
\authornote{Corresponding author.}
\author{Rongrong Ji}
\affiliation{%
  \department{Key Laboratory of Multimedia Trusted Perception and Efficient Computing, Ministry of Education of China}
  \country{School of Informatics, Xiamen University}
}
\email{{hui,rrji}@xmu.edu.cn}


\begin{abstract}
Refactoring is an indispensable practice of improving the quality and maintainability of source code in software evolution. Rename refactoring is the most frequently performed refactoring that suggests a new name for an identifier to enhance readability when the identifier is poorly named. However, most existing works only identify renaming activities between two versions of source code, while few works express concern about how to suggest a new name. In this paper, we study automatic rename refactoring on variable names, which is considered more challenging than other rename refactoring activities. We first point out the connections between rename refactoring and various prevalent learning paradigms and the difference between rename refactoring and general text generation in natural language processing. Based on our observations, we propose \ours, a two-stage pre-trained framework for rename refactoring on variable names. \ours first predicts the number of sub-tokens in the new name and then generates sub-tokens accordingly. Several techniques, including constrained masked language modeling, contrastive learning, and the bag-of-tokens loss, are incorporated into \ours to tailor it for automatic rename refactoring on variable names. Through extensive experiments on our constructed refactoring datasets, we show that the generated variable names of \ours are more accurate and meaningful than those produced by the existing method. Our implementation and data are available at \url{https://github.com/KDEGroup/RefBERT}.
\end{abstract}



\maketitle


\section{Introduction}
\label{sec:introduction}

Refactoring improves the internal code structure without altering external behavior~\cite{0019908}. 
It is a crucial activity often involved in software evolution, aiming at improving the quality of software projects and accelerating the process of development. 
Renaming refactoring identifies the inappropriate use of an identifier (e.g., variable, type, method and class names) and provides a meaningful name to replace. 
In practice, a large portion of source tokens are identifiers.
For example, Dei{\ss}enb{\"{o}}ck and Pizka~\cite{DeisenbockP05} analyze the code base of Eclipse and find that identifiers account for 32.8\% of the tokens and 71.7\% of the characters in the source code.
Developers often apply renaming operations for better readability and maintainability of source code, which helps avoid software fault-proneness~\cite{ArnaoudovaEPOAG14}, and renaming refactoring is the most frequently performed refactoring type that occupies developers' working time~\cite{MurphyKF06,GolubevKABM21,Murphy-HillPB12}.
Choosing a meaningful name for an identifier is non-trivial and demands both professional and contextual knowledge of programs.

In the literature, various works study refactoring, but most of them focus on automatic refactoring detection~\cite{DigCMJ06,PreteRSK10,KimGLR10,SilvaV17,TsantalisMEMD18,TsantalisKD22}, i.e., detect the refactoring activities between two versions of source code and identify the refactoring type.
For example, Malpohl et al.~\cite{MalpohlHT03} design a method to detect rename refactoring and Arnaoudova et al.~\cite{ArnaoudovaEPOAG14} propose REPENT to detect and classify identifier renaming. 
But these efforts aim at identifying whether refactoring operations exist between two versions of source code, rather than providing automatic renaming. 
Only a few works study automatic rename refactoring~\cite{ThiesR10,AllamanisBBS14,0008SMOBL17}.  
A representative work is \textsc{Naturalize}~\cite{AllamanisBBS14} which leverages the n-gram language model to estimate the probability that a specific name should be used in a given context to rename an identifier.

To accelerate rename refactoring and reduce the intellectual burden of developers, in this paper, we propose a two-stage pre-trained framework for automatic \underline{Re}name re\underline{F}actoring based on the \underline{BERT} architecture~\cite{DevlinCLT19} (\ours). 
Particularly, \textbf{we focus on rename refactoring on variable names}, which is considered much more challenging than refactoring other types of identifiers such as method and type names~\cite{AllamanisBBS14}.
\textbf{Consequently, throughout this paper, rename refactoring refers to rename refactoring on variable names.}
It is worthy noting that we refactor all the references of a variable name in one function.
There are other works for rename refactoring on other identifiers. 
For example, rename refactoring on method names~\cite{Liu0BKKKKT19,ParsaNER23,LiuLFLHJ22}. But they differ from our task. One variable name can appear in multiple positions in a function, and the positions of variables vary in different functions. Differently, method names typically appear in the beginning of the function.

The design of \ours is based on the following observations:
\begin{itemize}  
\item Rename refactoring is essentially similar to masked language modelling (MLM), which is a pretext task commonly used in pre-training BERT~\cite{DevlinCLT19,abs-1907-11692}. MLM fills the masked part of a text according to its context. Similarly, rename refactoring aims to suggest a meaningful variable name according to the context. Therefore, we believe MLM can be adopted for training an automatic rename refactoring model.

\item Unlike the variable name prediction task~\cite{RaychevVK19,VasilescuCD17,0002ZLY18}, both the context of the target variable and the variable name before refactoring are known in rename refactoring. Contrastive learning~\cite{LiuZHMWZT23}, which contrasts positive and negative samples for improving representation learning, is an ideal learning paradigm for automatic rename refactoring: we expect the generated name to be close to the variable name after refactoring but far away from the variable name before refactoring. 

\item Although rename refactoring can be viewed as a generation task in natural language processing (NLP), the standard cross-entropy loss for text generation is not suitable for rename refactoring. Unlike natural language text where words should follow a strict order to ensure grammatical correctness, sub-tokens in a variable name do not have such a restriction. Different orders of sub-tokens for a variable name do not significantly affect our understanding of the variable. Thus, the standard cross-entropy loss that emphasizes the strict alignment between the prediction and the target is suboptimal for automatic rename refactoring.

\end{itemize}

Based on the above observations, we propose various techniques to endow \ours with the ability of renaming variables.
In summary, our contributions are:
\begin{itemize} 

\item We point out the connections between rename refactoring and various prevalent learning paradigms (MLM and contrastive learning) and the difference between rename refactoring and general text generation in NLP (the inappropriateness of using standard cross-entropy loss), which, as far as we know, has not been mentioned in the literature. 

\item We construct a refactoring dataset \textsc{JavaRef}, and it can be used to evaluate automatic rename refactoring and other refactoring tasks. We further adapt an existing code corpus \textsc{TL-CodeSum} as a supplement to \textsc{JavaRef}.

\item We design a two-stage framework \ours for automatic rename refactoring. 
\ours first predicts the number of sub-tokens in the new name and then generates sub-tokens accordingly.
Various techniques, including constrained masked language modeling, contrastive learning, and the bag-of-tokens loss, are incorporated into \ours to tailor it for automatic rename refactoring.

\item We conduct extensive experiments to demonstrate the effectiveness of \ours. The experimental results show that \ours provides more accurate and meaningful suggestions for rename refactoring than the existing method.

\end{itemize}

The rest of this paper is organized as follows.
In Sec.~\ref{sec:relatedwork}, we introduce the related work of \ours.
In Sec.~\ref{sec:data}, we describe how we collect and construct two refactoring datasets \textsc{JavaRef} and \textsc{TL-CodeSum}.
Our framework \ours is illustrated in Sec.~\ref{sec:method}.
We then demonstrate experimental settings in Sec.~\ref{sec:exp_set}. 
The experimental results are reported and analyzed in Sec.~\ref{sec:exp}.
In Sec.~\ref{sec:threats}, we discuss possible threats to the validity of our work.
Finally, we conclude our work in Sec.~\ref{sec:con}.


\section{Related Work}
\label{sec:relatedwork}

In this section, we discuss several areas that are related to our work.

\subsection{Automatic Refactoring Detection}

Existing studies of refactoring mostly focus on detecting refactoring activities between two versions of source code and identifying the refactoring type, i.e., automatic refactoring detection.
Automatic refactoring detection helps update applications to use the new version of its components that have changed the interfaces~\cite{DigCMJ06}, and enhance the understanding of software evolution~\cite{RatzingerSG08,SilvaTV16}. 
Wei{\ss}gerber and Diehl~\cite{WeissgerberD06} identify refactorings based on clone detection.
Demeyer et al.~\cite{DemeyerDN00} propose a set of heuristics based on object-oriented metrics for automatic refactoring detection.
Negara et al.~\cite{NegaraCVJD13} design Refactoring Inference Algorithm that can detect 10 refactorings according to refactoring properties.
Dig et al.~\cite{DigCMJ06} develop \textsc{RefactoringCrawler} that applies syntactic analysis and semantic analysis to detect 7 types of refactorings in Java projects. 
Prete et al.~\cite{PreteRSK10} and Kim et al.~\cite{KimGLR10} design \textsc{Ref-Finder} which expresses each refactoring type in terms of template logic rules and uses a logic programming engine to detect refactorings.
Tsantalis et al.~\cite{TsantalisMEMD18,TsantalisKD22} propose \textsc{RefactoringMiner} that uses an abstract syntax tree (AST) based statement matching algorithm to detect refactorings.
Silva and Valente~\cite{SilvaV17} propose \textsc{RefDiff}, which employs both heuristics based on static analysis and code similarity to detect 13 refactoring types. 
There are also a few works specially designed for detecting the rename refactoring~\cite{MalpohlHT03,ArnaoudovaEPOAG14}. 

Unlike the above works that take two versions of source code as inputs for detecting refactorings, our work requires a method containing a target variable as input and provides suggestions for rename refactoring, i.e., variable rename refactoring.

\subsection{Automatic Rename Refactoring}
\label{sec:rr}

Developers spent much of their time in software refactoring~\cite{GolubevKABM21} and rename refactoring is the most frequently performed refactoring operations, as reported by Murphy-Hill et al.~\cite{Murphy-HillPB12}. 

There are a few works on automatic rename refactoring\footnote{Note that, in Sec.~\ref{sec:rr}, we introduce related work of general rename refactoring on different types of identifiers.
But this paper focus on rename refactoring on variable names.}.
Caprile and Tonella~\cite{CaprileT00} first map each word in the identifier to a standard lexicon and then the sequence of words for the identifier is required to be compliant with a grammar.
Thies and Roth~\cite{ThiesR10} present a tool to support identifier renaming based on information extracted via static code analysis.
Feldthaus and M{\o}ller~\cite{FeldthausM13} propose a semi-automatic refactoring method that combines static analysis and interaction with the programmer for rename refactoring in JavaScript. 
Mayer and Schroeder~\cite{MayerS14} apply multiple existing language-specific refactoring routines for multi-language artifact binding and rename refactoring.
\textsc{Naturalize}, developed by Allamanis et al.~\cite{AllamanisBBS14}, is the prior work that adopts NLP techniques for refactoring. \textsc{Naturalize} leverages a n-gram language model to suggest identifier names to replace those that break coding conventions. It first retrieves a set of candidate names that have appeared in the similar context from other code snippets, and then ranks the candidates by measuring naturalness through a learned n-gram model. 
The idea of \textsc{Naturalize} has inspired other researchers to explore using language modeling for rename refactoring~\cite{0008SMOBL17}.

Most of the above works are rule-based or semi-automatic while \ours is a learning-based rename refactoring method that does not require complex features and rules.
\textsc{Naturalize} that adopts language modeling for automatic rename refactoring is mostly related to our work. 
However, \ours differs from \textsc{Naturalize} in that we employ a pre-trained framework that benefits rename refactoring by leveraging a large volume of public code data. 
Besides, \ours conducts token-level rename refactoring that better captures the semantics of variables.

\subsection{Variable Name Prediction}
\label{sec:var_predict}
Variable name prediction is a related but different task to rename refactoring.
The goal of variable name prediction is to generate a variable name that is as close as possible to the missing variable name.
Differently, rename refactoring generates variable names from the perspective of code refactoring and the resulting variable name may be quite different but more readable compared to the original one.

There are some studies on variable name prediction.
Raychev et al.~\cite{RaychevVK19} leverage conditional random fields to predict variable names by modeling relations among variables and program elements.
Vasilescu et al.~\cite{VasilescuCD17} recover variable names through statistical machine translation. 
Alon et al.~\cite{0002ZLY18} represent a program using paths in its AST to predict method names, variable names and variable types. 

\subsection{Code Representation Learning}
Recently, learning code representations has attracted great attention since it can assist various code-related tasks, including but not limited to code summarization~\cite{LinOZCLW21}, code completion~\cite{WangL21a} and code search~\cite{LeCB20}.
A great number of works on code representation learning have sprung up.
Earlier approaches apply the idea of word2vec~\cite{MikolovSCCD13} to obtain code representations~\cite{BojanowskiGJM17, abs-1904-03061}.
More recent works~\cite{abs-2004-13214, abs-2001-00059, FengGTDFGS0LJZ20} mostly adopt pre-training techniques which can fully leverage the large volume of public source code in platforms like GitHub. 
Most of them are built based on the BERT architecture~\cite{DevlinCLT19}, which has achieved dramatic empirical improvements in multiple AI tasks. 
Some representative works include CodeBERT~\cite{FengGTDFGS0LJZ20}, GraphCodeBert~\cite{GuoRLFT0ZDSFTDC21}, T5~\cite{MastropaoloSCNP21},  UniXcoder~\cite{GuoLDW0022} and SPT-code~\cite{NiuL0GH022}, and they have benefited various downstream code-related tasks.


\section{Data Construction}
\label{sec:data}

We can utilize existing large code corpora, even those not related to rename refactoring, to pre-trained \ours to enhance its general understanding of programming language.
When fine-tuning \ours on automatic rename refactoring, \ours requires rename refactoring data corpora that can be smaller than pre-training data but contain ground-truth rename refactoring.
To prepare datasets for training and evaluating automatic rename refactoring approaches, we use rules to modify an existing code corpus \textsc{TL-CodeSum}\footnote{\url{https://github.com/xing-hu/TL-CodeSum}}~\cite{HuLXLJ20} originally designed for the code summarization task so that it can be used in the automatic rename refactoring task. Additionally, we collect and construct a Java refactoring corpus named \textsc{JavaRef} through manual inspection and using automated tools.

\subsection{Adapt Existing Code Corpus for Rename Refactoring}

\begin{figure}[t]
    \centering
    \subfloat{\includegraphics[width=0.46\columnwidth]{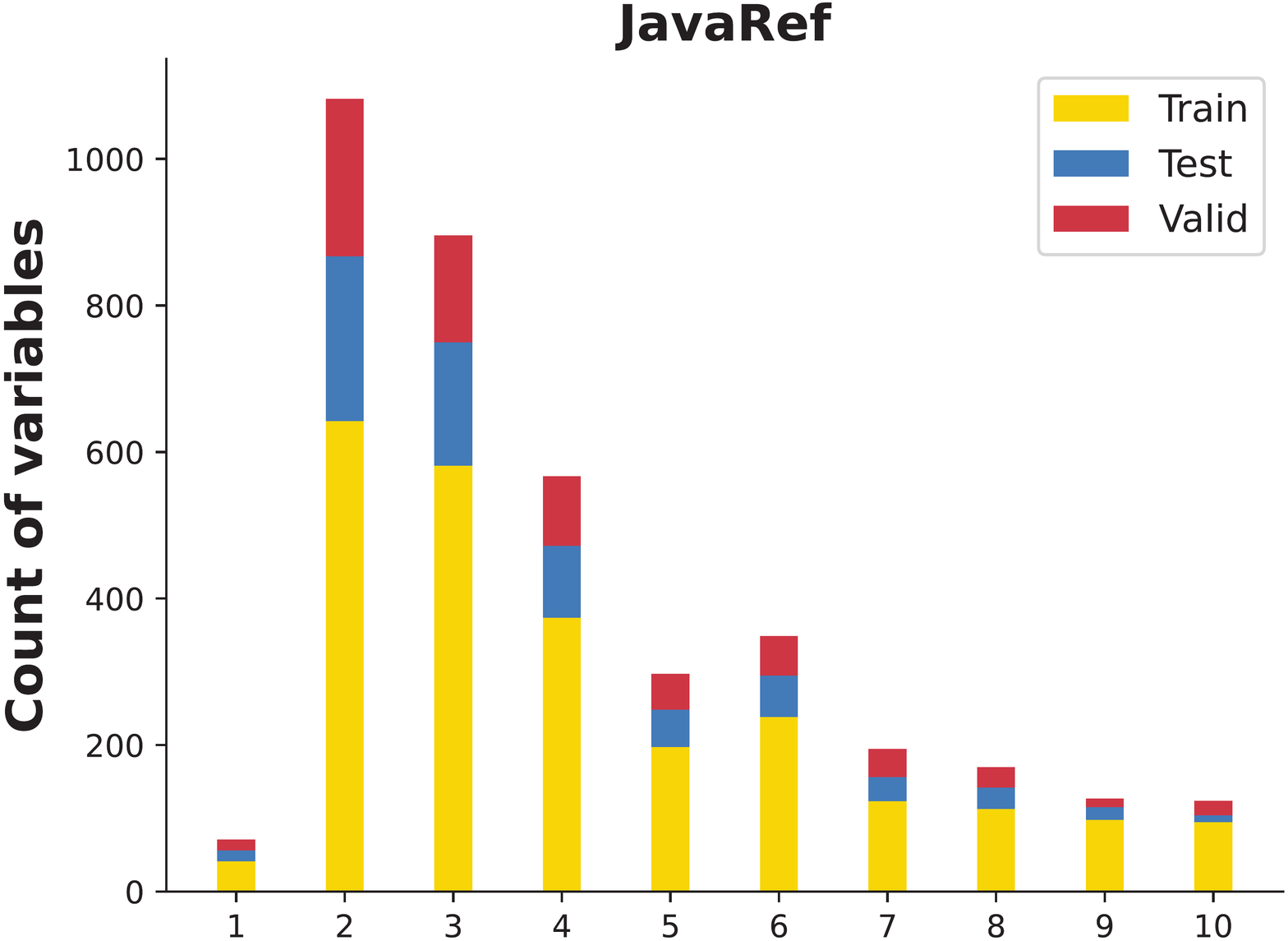}}
    \qquad
    \subfloat{\includegraphics[width=0.46\columnwidth]{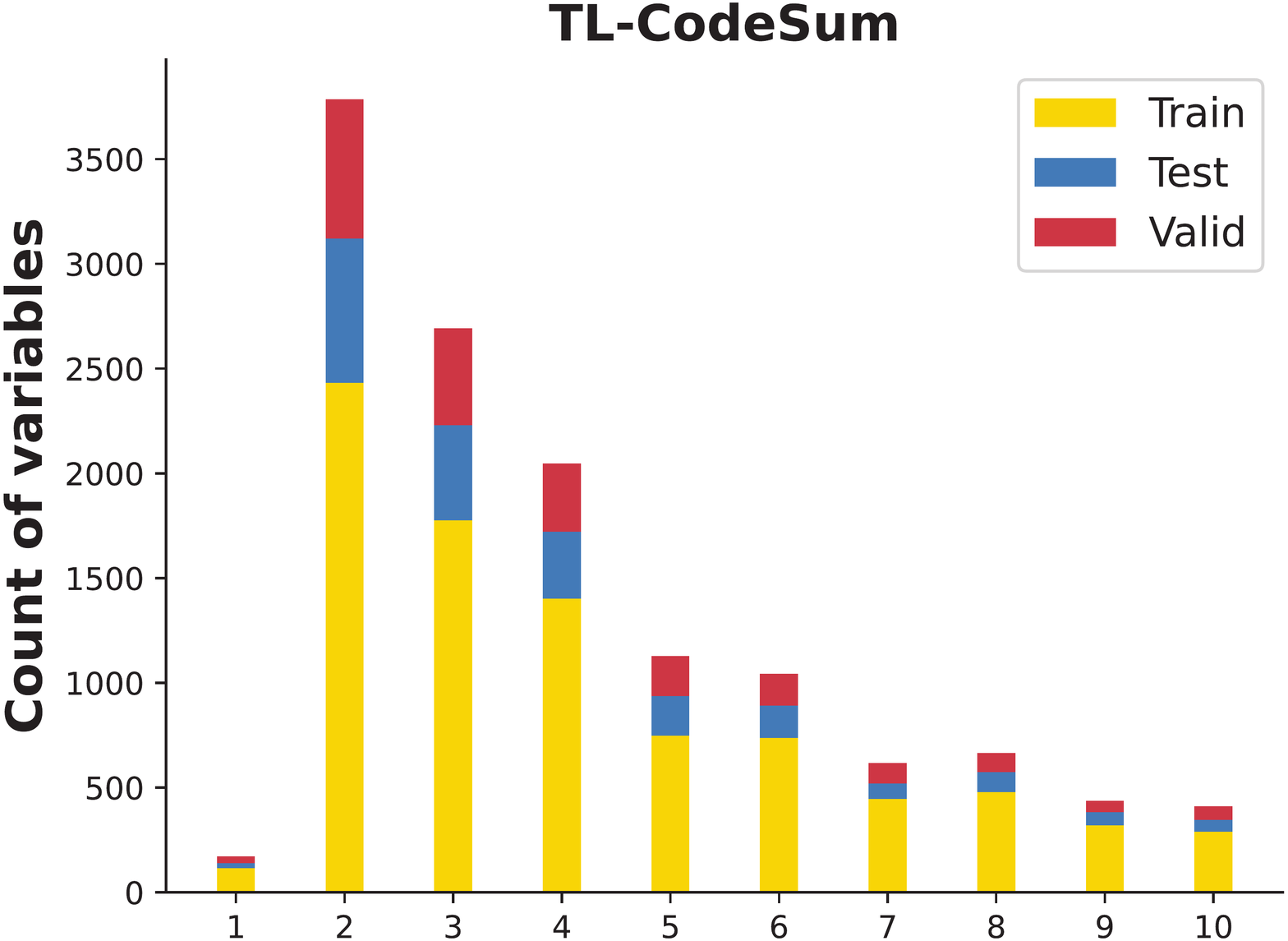}}
    \caption{Distributions of the frequency of variable names.}
    \label{fig:var_fre}
\end{figure}

Since the cost of manually labeling a large refactoring corpus is exorbitant, we first adapt the existing code corpus for the automatic rename refactoring task.
We choose \textsc{TL-CodeSum} from the literature as the target dataset.
\textsc{TL-CodeSum} is originally used in the code summarization task. 
We use \textsc{AstParser}\footnote{\url{https://github.com/Ragnaroek/eclipse-astparser}} to parse a function into an AST and retrieve variable names from AST nodes. 
Since the original \textsc{TL-CodeSum} does not have code before/after rename refactoring, we assume that variable names in \textsc{TL-CodeSum} are already meaningfully and reasonably designed, and they do not require rename refactoring.
We perform the following steps to adapt \textsc{TL-CodeSum}:
\begin{enumerate}
	\item We randomly pick one variable name from each function in \textsc{TL-CodeSum} to construct a variable name set.

	\item For each function, the picked variable name $v$ is treated as the name after rename refactoring (i.e., ground-truth refactored variable names). We further sample another variable name $v'$ from the variable name set as the name of $v$ before rename refactoring.

	\item We perform subword tokenization on the modified \textsc{TL-CodeSum}.  
	We adopt the subword tokenization approach of Byte-Pair Encoding~\cite{SennrichHB16a} used by RoBERTa-base~\cite{abs-1907-11692} to split code into subwords and variable names are divided into tokens with a finer granularity. This step is essential for rename refactoring. As the data of variable names is extremely sparse (i.e., developers can name a variable freely and most variable names only appear a few times in a large code corpus as shown in Fig.~\ref{fig:var_fre}), variable names' sub-tokens that are more likely to occur before are a better encoding unit. Encoding sub-tokens, which occur more frequently, can better capture the semantics of variable names and alleviate the data sparsity issue of variable names.
\end{enumerate}

Finally, we have a large dataset for automatic rename refactoring and it contains 65,324 rename refactoring data records.

\subsection{Construction of \textsc{JavaRef}} 
The construction of the Java refactoring corpus \textsc{JavaRef} involves two parts:

\vspace{5pt}
\noindent\textbf{Data Collection.} 
This part contains several data collection steps:
\begin{enumerate} 
	\item Tsantalis et al.~\cite{TsantalisMEMD18} and Claes et al.~\cite{ClaesM20} provide two datasets containing commit data in many open-source Java projects. Based on the project popularity (i.e., numbers of stars and forks), we select 114 java open-source GitHub projects containing 8,451 commits from their datasets. 
 
	\item Then, to extract the refactoring history from commit data, we employ a popular refactoring type detector \textsc{RefactoringMiner}~\cite{TsantalisMEMD18,TsantalisKD22} which shows high refactoring detection accuracy. The detection result contains the hash value of a commit. It also includes refactoring types and corresponding descriptions between two versions of the project before and after the commit. But the result does not have the information of changed source code files and code diff data.

	\item Based on the retrieved refactoring information, we further adopt a Python library \textsc{PyDriller}\footnote{\url{https://github.com/ishepard/pydriller}} to extract the url of each changed source code file and line numbers of the changed code. We crawl changed source code files before/after refactoring from GitHub according to these urls.

\end{enumerate}
 
\noindent\textbf{Data Preprocessing.} 
We perform the following preprocessing steps in sequence:
\begin{enumerate} 
	\item To improve data quality and eliminate the noise brought by the incorrect results from \textsc{RefactoringMiner} and \textsc{PyDriller}, we invite five students in Computer Science major to manually verify the collected data in a period of three months. They are asked to link a changed code snippet (i.e., code diff) to one of the refactoring types detected by using \textsc{RefactoringMiner} to compare the two commits where the code snippet is changed.

	\item Then, we keep verified, function-level refactoring data and filter other refactorings. We further remove duplicated code and functions shorter than three lines.  

	\item We further extract the variable names in the data record. We use \textsc{AstParser} to parse the function into an AST and retrieve variable names from AST nodes.
 
	\item Finally, we perform subword tokenization on \textsc{JavaRef} in a similar way as preprocessing \textsc{TL-CodeSum}.
\end{enumerate}

In total, \textsc{JavaRef} contains 17,910 refactorings, covering 25 refactoring types.  
Each data record in \textsc{JavaRef} consists of refactoring type, refactoring description, original code (before refactoring) and current code (after refactoring). 
If a code sample in \textsc{JavaRef} does not correspond to a rename refactoring operation, we use the same method for preprocessing \textsc{TL-CodeSum} to prepare variable names before and after rename refactoring.


\section{Our Framework \ours}
\label{sec:method}
\subsection{Overview}

\begin{figure*}[t]
    \centering
    \includegraphics[width=0.7\linewidth]{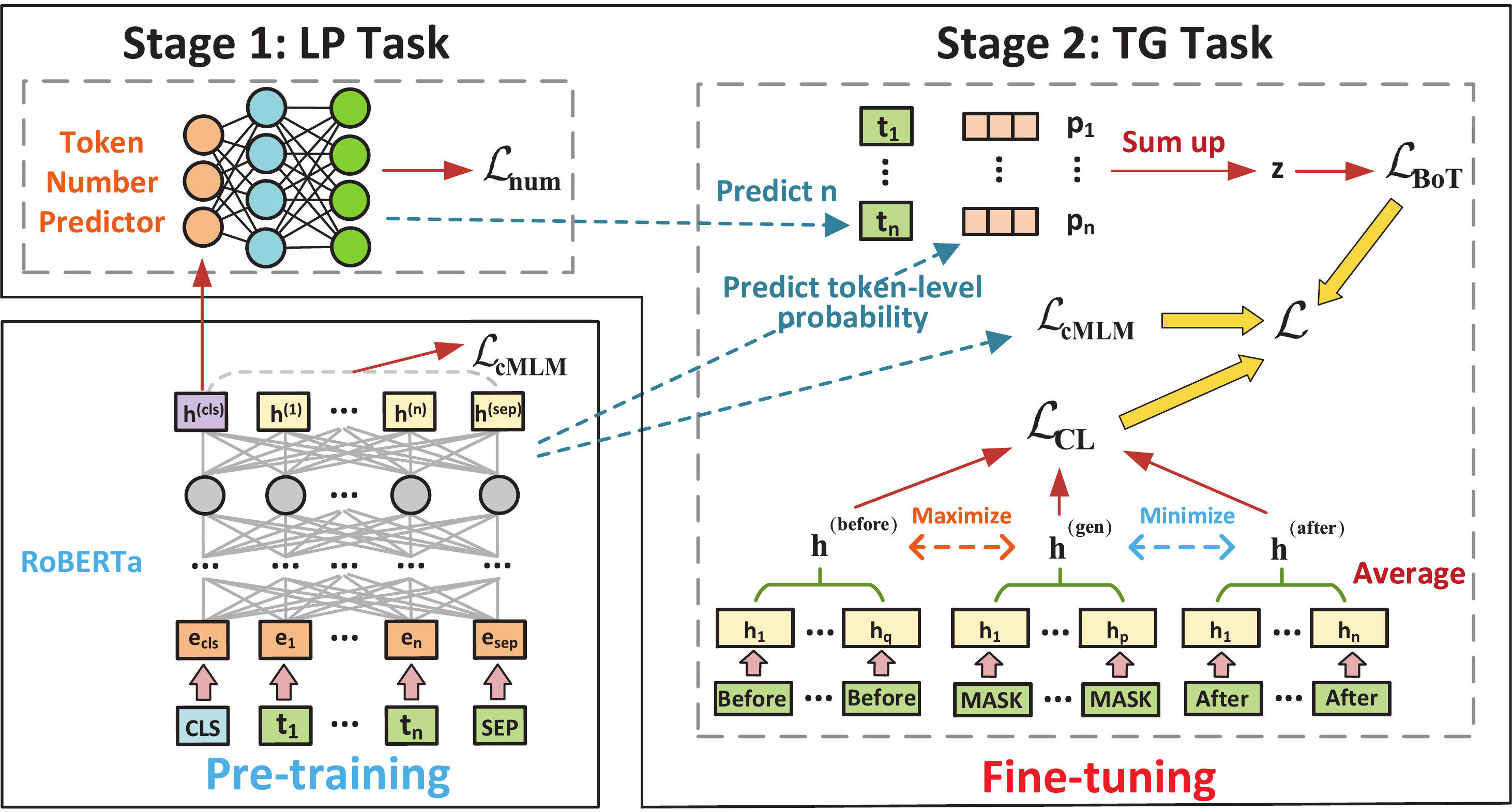}
    \caption{Overview of \ours.}
    \label{fig:overview}
\end{figure*}

\begin{figure}[t]
    \centering
    \includegraphics[width=0.85\columnwidth]{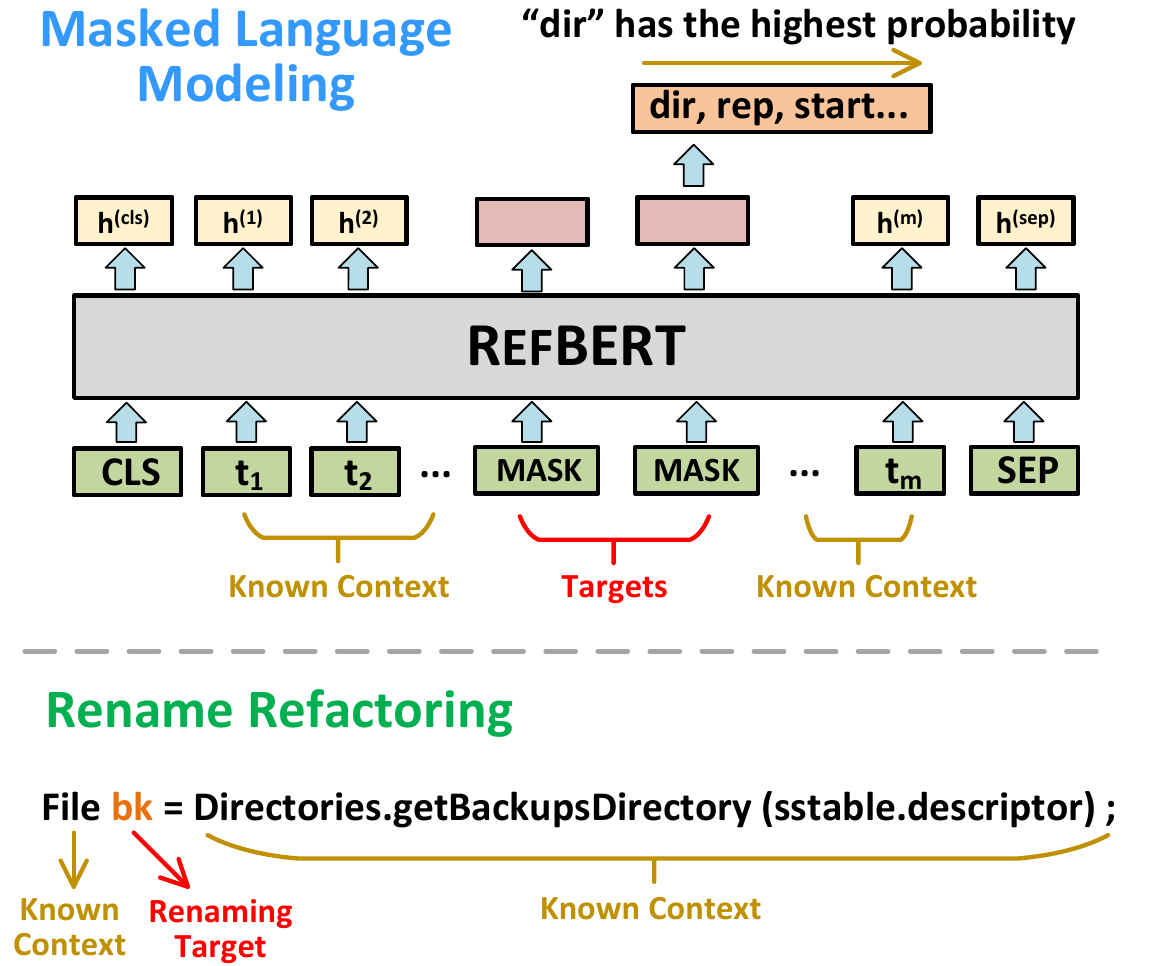}
    \caption{A comparison of MLM and rename refactoring.}
    \label{fig:mlm}
\end{figure}

Compared to general-purpose code corpora like \textsc{CodeSearchNet}~\cite{CodeSearchNet}, the volume of rename refactoring data corpora is relatively small.
Inspired by the success of large code pre-trained models~\cite{FengGTDFGS0LJZ20,GuoRLFT0ZDSFTDC21,MastropaoloSCNP21,GuoLDW0022,NiuL0GH022} on various downstream code-related tasks like code search and code summarization, we opt to adopt a pre-trained architecture for \ours.
In short, \ours is firstly pre-trained over large-scale code corpora and then fine-tuned over relatively smaller refactoring data.

Our design is based on not only the conclusion from previous studies~\cite{FengGTDFGS0LJZ20,NiuL0GH022} that pre-training can endow the model with a better generalization ability via training over large related data but also the following observations:  
(1) the prevalent pretext task in pre-training is essentially similar to rename refactoring; 
(2) the recently popular self-supervised learning paradigm, contrastive learning~\cite{LiuZHMWZT23}, fits the nature of rename refactoring and can be adopted in the fine-tuning phase;
(3) sub-tokens in a variable do not need to follow a strict order.

Fig.~\ref{fig:overview} provides an overview of \ours.
\ours is trained to generate refactored variable names in two steps: 
 
\vspace{5pt}
\noindent\textbf{Length Prediction (LP)}: In the LP task, \ours predicts the number of tokens in the refactored variable name. 

\vspace{5pt}
\noindent\textbf{Token Generation (TG)}: Given the predicted number of tokens, \ours generates tokens in the refactored variable name in the TG task.

\subsection{Architecture of \ours}
Prevalent pre-trained models in NLP typically adopt the architecture of BERT~\cite{DevlinCLT19} which contains a multi-layer bidirectional Transformer~\cite{VaswaniSPUJGKP17}. 
We follow this design and use 12 RoBERTa layers in \ours (i.e., RoBERTa-base in~\cite{abs-1907-11692}), which is a replication of the original BERT model with improved performance.
Given the widespread adoption of Transformer in software engineering field~\cite{YangXLG22}, we do not introduce its background in detail.

\vspace{5pt}
\noindent\textbf{Model Input.} \ours takes tokens of a code snippet as input: $\{\text{[CLS]}, t_1, \cdots, t_g, \text{[SEP]}\}$, where $t$ denotes a token and $g$ is the number of tokens in the code snippet. [CLS] and [SEP] are two special tokens indicating the start and end of a sequence.

\vspace{5pt}
\noindent\textbf{Model Output.} 
\ours produces two types of output: the contextual representation vector for each token and the representation of [CLS]. The latter serves as the aggregated sequence representation and is employed to predict the number of tokens in a variable name (see Sec.~\ref{sec:pred_number}).

\subsection{Pre-train \ours}

During pre-training, \ours can benefit from training over large code corpora and we adopt two datasets \textsc{CodeSearchNet}~\cite{CodeSearchNet} and \textsc{Java-small}~\cite{AlonBLY19} in pre-training.
Their details are described in Sec.~\ref{sec:exp_data}.

We find that rename refactoring is essentially similar to \textit{masked language modeling} (MLM), which is the primary pretext task used in various pre-trained models~\cite{FengGTDFGS0LJZ20, DevlinCLT19, JoshiCLWZL20, b27}.
As shown in Fig.~\ref{fig:mlm}, MLM masks some tokens in the input and trains the model to precisely predict masked tokens based on unmasked tokens. 
MLM does not require additional labels and can leverage intrinsic information of the input data.
Similarly, in rename refactoring, the goal is to predict the unknown tokens constituting the variable name based on other known tokens in the code snippet.
If we treat unknown and known tokens in rename refactoring as masked tokens and unmasked tokens in MLM, we can find that the two tasks are similar as depicted in Fig.~\ref{fig:mlm}.
MLM has been proven effective for learning rich token representations in various applications~\cite{DevlinCLT19, b26}.
The similarity between MLM and rename refactoring and the success of MLM in other domains have motivated us to adopt the idea of MLM in pre-training \ours for rename refactoring.

To be specific, the objective of MLM is to predict the original token, which is replaced by a [MASK] token during training, based only on its context. 
Let $T^{(m)}_{s}=\{t^{(m)}_{s,1},\cdots,t^{(m)}_{s,g^{(m)}_s} \}$ and $T^{(u)}_{s}=\{t^{(u)}_{s,1},\cdots,t^{(u)}_{s,g^{(u)}_s}\}$ denote the set of masked tokens and unmasked tokens in a code snippet $s$, respectively. 
$g^{(m)}_s$ and $g^{(u)}_s$ are the numbers of masked tokens and unmasked tokens, respectively.
The tokens in $T^{(m)}_{s}$ and $T^{(u)}_{s}$ are sorted according to their order in $s$. 
Then, the objective of MLM can be defined as:
\begin{equation}   
\label{eq:MLM}
\mathcal{L}_{\text{MLM}}(\theta)=-\sum_{s\in \mathcal{S}} \sum_{i=1}^{g^{(m)}_{s}}\log p\big( t^{(m)}_{s,i} \big| T^{(u)}_{s};\theta\big), 
\end{equation}
where $\theta$ indicates the model parameters and $\mathcal{S}$ is the code corpus. 

There exist various masking strategies for MLM.
The most common one is to randomly mask the tokens in sentences with a fixed ratio, e.g., 15\% in BERT \cite{DevlinCLT19}. 
Since rename refactoring focuses on variable names in a code snippet, we propose to use \emph{constrained masked language modeling} (cMLM) in \ours.
cMLM masks chosen tokens instead of randomly picked tokens.
Given a refactored code snippet $s$, \ours always masks all tokens in the target variable name with $g^{(m)}_s$ [MASK] tokens. 

The representation of the $i$-th [MASK] token ($1 \leq i\leq g_s^{(m)} $) from \ours is fed into an output softmax over the token vocabulary $\mathcal{V}$ to produce the probability distribution $\mathbf{p}_{s,i} \in \mathbb{R}^{\left| \mathcal{V} \right|}$ of all tokens for being the corresponding masked token, where $\left| \mathcal{V} \right|$ denotes the vocabulary size.
The cMLM task can be trained with a standard cross entropy loss for classification:
\begin{equation}   
\label{eq:cmlm}
\mathcal{L}_{\text{cMLM}}=-\sum_{s\in \mathcal{S}}\sum_{i=1}^{g_{s}^{(m)}}\sum_{j \in \mathcal{V}} y_{s,i,j} \log p_{s,i,j},
\end{equation}
where $y_{s,i,j}$ equals 1 if the $i$-th masked token in code snippet $s$ is token $j$ otherwise 0. 
$p_{s,i,j}$ is the predicted probability of the $i$-th masked token being $j$ (i.e., $i$-th dimension in $\mathbf{p}_{s,i}$) and it is produced by the output softmax.
 
It is worth pointing out that we adopt cMLM in both pre-training phase and fine-tuning phase since it is critical for training a rename refactoring model.

\subsection{Fine-tune \ours}

There are two stages in the fine-tuning phase. 
\ours is firstly trained to predict token number.
Then it is guided to generate variable name tokens precisely over refactoring data (\textsc{JavaRef} and \textsc{TL-CodeSum}).
\subsubsection{Stage 1: Predict Token Number (LP task).}
\label{sec:pred_number}

Since code corpora have been preprocessed and a variable name is split into multiple tokens in order to better capture its semantics and alleviate the data sparsity issue,
before generating tokens, \ours should first predict the number of tokens (i.e., masked token number $g^{(m)}_{s}$) in the variable name after refactoring.

The representation of the first special token [CLS] in each input sequence to pre-trained models is commonly used for sentence classification or ranking tasks~\cite{DevlinCLT19,b26}. 
Inspired by this, we use the representation $\mathbf{h}^{(\text{cls})}_s$ of the [CLS] token in a code snippet $s$ to predict $g^{(m)}_s$. 
Note that, unlike cMLM which masks each token in a variable name with one [MASK] token, \emph{we mask each variable name with only one [NUM] token in predicting the number of tokens} to avoid data leakage since the number of masked tokens reveals the ground truth. 
We feed $\mathbf{h}^{(\text{cls})}_s$ into a single-layer feedforward neural network (token number predictor) to predict the number of tokens in the masked variable name of $s$:
\begin{equation}
\label{eq:tok_num_pred}
\mathbf{q}_s = \mathbf{W}_{q}\mathbf{h}^{(\text{cls})}_s + \mathbf{b}_{q},
\end{equation}
where $\mathbf{W}\in \mathbb{R}^{l_{\text{max}}\times d}$ and $\mathbf{b}_{l} \in \mathbb{R}^{l_{\text{max}}}$ are learnable parameters.
$l_{\text{max}}$ is the pre-defined maximum number of tokens in a variable name, which is derived from a statistical analysis of the different variable names present in the training dataset.
$d$ indicates the dimensionality of representations.
$\mathbf{q}_s$ is the predicted probability distribution of the token number for the code snippet $s$.

We adopt a standard cross-entropy for predicting the number of tokens:
\begin{equation} 
\label{eq:len_obj}  
\mathcal{L}_{\text{num}}=-\sum_{s\in \mathcal{S}}\sum_{i=1}^{l_{\text{max}}} r_{s,i} \log q_{s,i},
\end{equation}
where $r_{s,i}$ equals 1 if the masked variable name in $s$ has $i$ tokens otherwise 0.
$q_{s,i}$ is the $i$-th dimension in $\mathbf{q}_{s,i}$, i.e., the predicted probability that the masked variable name has $i$ tokens.

\subsubsection{Stage 2: Generate Tokens (TG task).}
\label{sec:generate_token}

In Stage 2, \ours generates tokens in refactored variable names.

\vspace{5pt}
\noindent\textbf{Bag-of-token Loss for Guiding Token Generation.}
In many NLP applications, the generated sentence consisting of multiple words are expected to be similar to the target sentences in order to express the correct semantics. 
Word order deviations in sentences should yield excessively large losses since different word orders indicates different meanings or even violates the grammar rules of natural languages. 
Therefore, training a generative model with a cross-entropy loss is a standard scheme in NLP.
The cross-entropy loss is computed strictly between aligned ground-truth tokens and predicted tokens. 

However, in rename refactoring, the order of the generated tokens for the refactored variable name may vary.
Due to different coding conventions, it is possible that different developers use different orders of tokens to express the same variable name.
For example, for the ground-truth refactored variable name \textit{defaultVersion}, \textit{versionDefault} which contain same tokens \textit{default} and \textit{version} in a different order, expresses the same meaning.
And the different token orders do not violate the rule of programming languages since \textit{defaultVersion} and \textit{versionDefault} are processed integrally as variable names instead of individual tokens by the compiler.
However, if we solely adopt the cross-entropy loss to guide the generation of variable name tokens, \textit{versionDefault}  will be punished as the incorrect answer since [`version', `default'] is not strictly aligned with [`default', `version'], which potentially harms the training of the model. 

\begin{figure}[t]
\centering
\includegraphics[width=0.95\columnwidth]{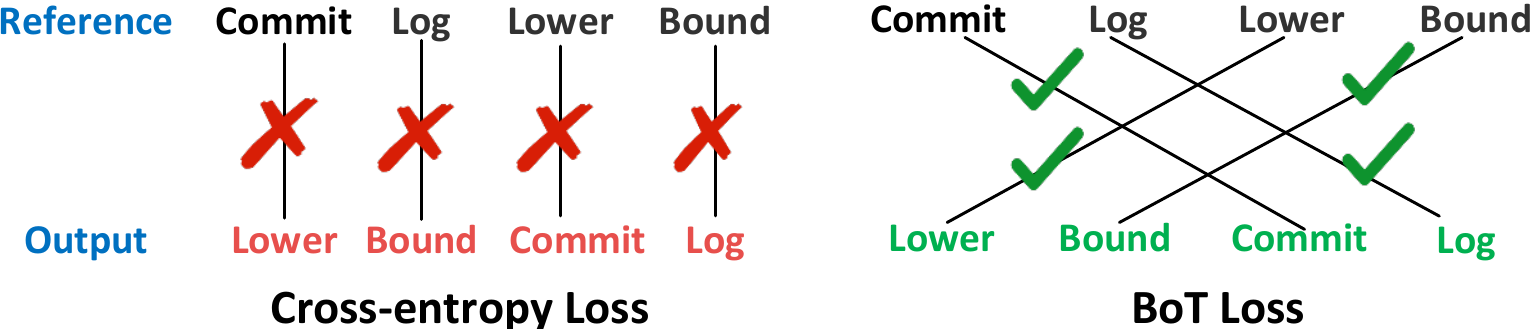} 
\caption{A comparison of cross-entropy loss and BoT loss.} 
\label{fig:loss_compare}
\end{figure}

Consequently, we consider loosening the strict restriction of token orders in the cross-entropy loss to guide the generation of variable name tokens by introducing a new loss. 
We adopt the idea of using bag-of-words (BoW)~\cite{ZhangJZ10} in designing text generation losses~\cite{MaSWL18,LiuCYW21} and modify the cross-entropy loss to a \emph{bag-of-tokens} (BoT) loss to reduce the punishment for token deviations in rename refactoring.
BoW is a text representation where a text is represented as the bag (multiset) of its words, disregarding grammar, word order and word frequency.
Similarly, when using BoT, we do not consider token order and token frequency in a variable name.
For instance, variable names \textit{CommitLogLowerBound} and \textit{LowerBoundCommitLog} have the same BoT representation \{\textit{Commit}, \textit{Log}, \textit{Lower}, \textit{Bound}\}.
Fig.~\ref{fig:loss_compare} illustrates the difference between the BoT loss and the cross-entropy loss.

The idea of the BoT loss is to compare the BoT representations of the generated variable names and the ground-truth, refactored variable name.
This way, token order deviation will not incur punishment of large losses as long as the generated variable names have similar tokens as the ground-truth, refactored variable name.
In the BoT loss, the probability distributions $\{ \mathbf{p}_{s,1}, \cdots, \mathbf{p}_{s,i} \}$ ($1 \leq i\leq g_{s}^{(m)}$) of [MASK] tokens predicted by \ours for all positions are summed up to form the variable-name-level distribution $\mathbf{z}_s \in \mathbb{R}^{\left| V \right|}$ for the code snippet $s$:
\begin{equation}
\label{eq:variable_name_pro}
\mathbf{z}_s =  \sum_{i=1}^{g_{s}^{(m)}} sigmoid(\mathbf{p}_{s,i}),
\end{equation}
where $sigmoid(\cdot)$ indicates the dimension-wise calculation using the sigmoid function.
Each dimension in the variable-name-level distribution $\mathbf{z}_s$ represents how possible the corresponding token appears in the BoT for the refactored variable name in $s$ regardless of the token order.
Compared to the token-level probability distribution $\mathbf{p}_{s}$, the variable-name-level probability distribution $\mathbf{z}_s$ removes the restriction of the token order.
Based on $\mathbf{z}_s$, we can calculate the BoT loss which guides token generation as follows:
\begin{equation}   
\label{eq:bot}
\mathcal{L}_{\text{BoT}}=-\sum_{s\in \mathcal{S}}\sum_{i=1}^{g_{s}^{(m)}}\sum_{j \in \mathcal{V}} y_{s,i,j} \log z_{s,j},
\end{equation}
where $z_{s,j}$ is the predicted probability of the $i$-th masked token being $j$ (i.e., $i$-th dimension in $\mathbf{z}_{s}$) and other notations have same meanings as Eq.~\ref{eq:cmlm}.

\vspace{5pt}
\noindent\textbf{Contrast Variable Names Before and After Refactoring.} 
As discussed in Sec.~\ref{sec:var_predict}, rename refactoring is different from the variable name prediction task since the variable name after refactoring is quite different compared to the name before refactoring.
To fully utilize such a difference to improve rename refactoring, we opt for contrastive learning, a type of self-supervised learning that has recently gained significant momentum~\cite{LiuZHMWZT23}.
Conceptually, contrastive learning aims to learn a representation space by minimizing the distance between positive samples while maximizing the distance between negative samples. 
A suitable contrast task will facilitate capturing intrinsic features in the data without requiring additional labels~\cite{LiuZHMWZT23}.

For rename refactoring, in the context of contrastive learning, the representation of the variable name after refactoring should be as close as possible to the representation of the ground-truth refactored variable name (i.e., positive sample), and far away from the variable name before refactoring (i.e., negative sample), as depicted in the bottom-right part of Fig.~\ref{fig:overview}.
We perform the average pooling operation on the representations of all tokens in a variable name $v$ to construct the representation $\hat{\mathbf{h}}_v$ of $v$:
\begin{equation}
\begin{aligned}
\mathbf{h}_v &= \text{Avg}\big(\{\mathbf{h}^{(1)},\cdots, \mathbf{h}^{(j)} \}\big), \,\, 1 \leq j \leq \left| v \right| \\
\hat{\mathbf{h}}_v &= \mathbf{h}_v / \left \| \mathbf{h}_v \right \|_{2}
\end{aligned}
\end{equation}
where $\left| v \right|$ is the number of tokens in $v$, $\mathbf{h}^{(1)}$ is the representation of the $j$-th token in $v$, $\text{Avg}(\cdot)$ is the average pooling operation, and $\left \| \cdot   \right \|_2$ denotes L2 normalization.

We adopt the non-parametric instance discrimination loss~\cite{WuXYL18} in \ours to achieve the goal of contrastive learning:
\begin{equation}  
\label{eq:contrast_loss}
\mathcal{L}_{CL}=-\sum_{i\in \mathcal{I}}\log\frac{e^{sim(\hat{\mathbf{h}}_{i}^{\text{(gen)}},\,\hat{\mathbf{h}}_{i}^{\text{(after)}} )/\tau}}{ e^{sim(\hat{\mathbf{h}}_{i}^{\text{(gen)}},\,\hat{\mathbf{h}}_{i}^{\text{(after)}} )/\tau} + e^{sim(\hat{\mathbf{h}}_{i}^{\text{(gen)}},\,\hat{\mathbf{h}}_{i}^{\text{(before)}} )/\tau}}, 
\end{equation}
where $\mathcal{I}$ is the refactored variable name set, $sim(\cdot )$ denotes the cosine similarity and $\tau$ is the temperature hyper-parameter that affects distribution concentration~\cite{WuXYL18}.
$\hat{\mathbf{h}}^{\text{(gen)}}$, $\hat{\mathbf{h}}^{\text{(before)}}$ and $\hat{\mathbf{h}}^{\text{(after)}}$ are representations of the predicted variable name, the variable name before refactoring and the variable name after refactoring, respectively.

\vspace{5pt}
\noindent\textbf{Complete Loss in Stage 2.}
The complete training loss in Stage 2 can be defined as follows:
\begin{equation} 
\label{eq:fine-tune-loss}  
\mathcal{L}_{\text{fine-tune}} = \lambda_{\text{cMLM}} \cdot \mathcal{L}_{\text{cMLM}} + \lambda_{\text{BoT}} \cdot \mathcal{L}_{\text{BoT}} + \lambda_{\text{CL}} \cdot \mathcal{L}_{\text{CL}} ,
\end{equation}
where $\lambda_{\text{cMLM}}$, $\lambda_{\text{BoT}}$ and $\lambda_{\text{CL}}$ are hyper-parameters for loss weights.
To balance the strict order consistency brought by the original cross-entropy loss for cMLM (Eq.~\ref{eq:cmlm}) and the token order flexibility brought by the BoT loss (Eq.~\ref{eq:bot}),
during fine-tuning, $\mathcal{L}_{\text{cMLM}}$ is still kept.
Note that, since calculating the BoT loss brings additional overhead and the pre-training corpora are much larger than the fine-tuning corpora, we do not adopt $\mathcal{L}_{\text{BoT}}$ in the pre-training phase.

\subsection{Predict with \ours}
\label{sec:predict}

At predication, for a variable name in a code snippet $s$ to be refactored, we first use the token number predictor in \ours to predict the number of tokens in the new variable name after refactoring (i.e., $g_s^{(m)}$).
After that, for each of the $g_s^{(m)}$ tokens in the refactored variable name, \ours predicts the token probability distribution $\mathbf{p}$, and it select the token from the token vocabulary $\mathcal{V}$ with the largest probability as the generated token.
Then, all the generated tokens are concatenated as the predicted variable name. 
This way, \ours will not produce the same token multiple times for a variable name.
Since a variable name typically does not contain many tokens and these tokens are usually different, our method can generate more readable refactored variable names, as shown in our experiments.


\section{Experiment Settings}
\label{sec:exp_set}
This section illustrates the details of the experiment settings.

\subsection{Data} 
\label{sec:exp_data}
We choose two public datasets that are commonly used in code representation learning to pre-train \ours:
\begin{itemize}
\item \textbf{\textsc{CodeSearchNet}}\footnote{\url{https://github.com/github/CodeSearchNet}}~\cite{CodeSearchNet}: It contains about 6 million functions from open-source GitHub repositories spanning 6 programming languages and is widely used in various code relevant tasks including but not limited to code retrieval~\cite{GuCM21}, code completion~\cite{CiniselliCPPPB21} and code summarization~\cite{LinOZCLW21}. 
We use the Java dataset in \textsc{CodeSearchNet}. 
We use the data split in the dataset for training, validation and test.

\item \textbf{\textsc{code2seq}}\footnote{\url{https://github.com/tech-srl/code2seq}}~\cite{AlonBLY19}: It contains 11 large Java projects originally provided by Allamanis et al.~\cite{AllamanisPS16}. We adopt its preprocessed version, \textsc{Java-small}~\cite{AlonBLY19}, in our experiments. It contains about 700K examples. We use the data split in the dataset for training, validation and test.
\end{itemize}

For the fine-tuning phase, our constructed rename refactoring datasets \textsc{JavaRef} and \textsc{TL-CodeSum} illustrated in Sec.~\ref{sec:data} are used on the rename refactoring task.

Tab.~\ref{tab:data} provides the statistics of all the datasets in the experiment.

\begin{center}
\begin{table}[t]
\caption{Statistics of datasets.}
\label{tab:data}
\scalebox{0.95}{
\begin{tabular}{c|c|c|c|c}
\hline
Stage                         & Dataset                & Training & Test   & Validation \\ \hline
\multirow{2}{*}{Pre-training} & \textsc{CodeSearchNet} & 410,050  & 24,607 & 11,004     \\
                              & \textsc{Code2Seq}      & 418,438  & 50,480 & 50,993     \\ \hline
\multirow{2}{*}{Fine-tuning}  & \textsc{JavaRef}       & 14,319   & 1,801  & 1,790      \\
                              & \textsc{TL-CodeSum}    & 50,908   & 7,243  & 7,173      \\ \hline
\end{tabular}
}
\end{table}
\end{center}

\subsection{Environment} We implement \ours\footnote{Our implementation and data are available at \url{https://github.com/KDEGroup/RefBERT}.} using PyTorch and run the experiments on a machine with two Intel(R) Xeon(R) Silver 4214R CPU @ 2.40GHz, 256 GB main memory and one NVIDIA GeForce RTX 3090.

\subsection{Evaluation}
\ours first predicts the number of tokens in the refactored variable name (the LP task), then it generates tokens in the refactored variable name accordingly (the TG task).
Hence, our evaluations involve two parts:

\subsubsection{Length Prediction (LP)}
The LP task corresponds to the first stage illustrated in Sec.~\ref{sec:pred_number}.

\vspace{5pt}
\noindent\textbf{Baseline}: We adopt a heuristic-based method as the baseline for the LP task. It selected the predicted number with the largest average probability from \ours, i.e., maximum $\frac{1}{g^{(m)}_{s}}\sum_{i=1}^{g^{(m)}_{s}}\log p\big( t^{(m)}_{s,i} \big| T^{(u)}_{s};\theta\big)$. 

\vspace{5pt}
\noindent\textbf{Metrics}: We choose Hit@$K$ as the evaluation metric. We evaluate whether the model predicts the ground-truth number as the most possible number (Hit@$1$) or among the top-$3$ most possible numbers (Hit@$3$)~\cite{GhazvininejadLL19}.

\subsubsection{Token Generation (TG)} 
The TG task corresponds to the second stage illustrated in Sec.~\ref{sec:generate_token}.

\vspace{5pt}
\noindent\textbf{Baseline}: \textsc{Naturalize}~\cite{AllamanisBBS14} is the baseline used in the TG task. It leverages a n-gram language model to suggest identifier names, including variable names, to replace those that break coding conventions. It first retrieves a set of candidate names that have appeared in the similar context from other code snippets, and then ranks the candidates by measuring naturalness through a learned n-gram model.  

\vspace{5pt}
\noindent\textbf{Metrics}: We adopt various metrics for the TG task.

\begin{itemize}
\item \textbf{Accuracy}: It measures the percentage of tokens, which are contained in the ground-truth variable name, can be generated by the model in the prediction. Large accuracy indicates good results.

\item \textbf{Exact Match (EM)}~\cite{AlonZLY19}\textbf{:} It measures the percentage of refactored variable names that can be exactly generated by the model. It is a variable-name-level metric and does not count the token-level similarity. For instance, the EM between ``ImageFolder'' and ``ImgFolder'' is 0. Large ED shows good results.
	        
\item \textbf{Edit Distance (ED)}: It compares the similarity of two strings by measuring the minimum number of editing operations w.r.t. tokens required to convert one string into another. For instance, the ED between ``ImageFolder'' and ``ImgFolder'' is 1. Small ED indicates good results.

\item \textbf{Character Error Rate (CER)}~\cite{b25}\textbf{:} It calculates the ED between the ground-truth variable name and the generated variable name. Then the result is normalized by the number of characters in the ground-truth variable name. 
Small CER indicates good results.

\end{itemize}

\subsection{Hyper-parameters}
We preserve the first 512 tokens in each code snippet at most.
Following RoBERTa~\cite{b26}, for \ours, we sample initial weights from $\mathcal N(0, 0.02)$, initialize biases to zero, set dropout ratio to 0.1 for all layers, and use GELU~\cite{hendrycks2016gaussian} as activation function in all layers. 
For predicting the token number, $l_{\text{max}}$ is set to 5 by default.
We conduct grid search for $\lambda$ and $\tau$ on the validation set.
The default $\lambda_{\text{CL}}$, $\lambda_{\text{cMLM}}$ and $\lambda_{\text{BoT}}$ are 1, 1 and 0.1, respectively.
The default $\tau$ is set to 0.05.
We optimize \ours using Adam \cite{KingmaB14}.
We terminate training when the method converges.

\begin{center}
\begin{table}[t]
\caption{Performance of the LP task.}
\label{tab:lp}
\scalebox{0.86}{
\begin{tabular}{c|c|cccc}
\hline
\multirow{2}{*}{Method}                                                           & \multirow{2}{*}{Dataset} & \multicolumn{2}{c}{\textsc{JavaRef}} & \multicolumn{2}{c}{\textsc{TL-CodeSum}} \\ \cline{3-6} 
    &                & Hit@1             & Hit@3            & Hit@1              & Hit@3              \\ \hline
\multirow{2}{*}{\begin{tabular}[c]{@{}c@{}}Heuristic-based\\ Method\end{tabular}} & \textsc{CodeSearchNet}   & 0.527             & 0.817            & 0.543              & 0.828              \\                                             & \textsc{Code2Seq}        & 0.506             & 0.782            & 0.529              & 0.792              \\ \hline
\multirow{2}{*}{\ours}                                                          & \textsc{CodeSearchNet}   & \textbf{0.655}    & \textbf{0.946}   & \textbf{0.702}     & \textbf{0.963}     \\           & \textsc{Code2Seq}        & \textbf{0.675}    & \textbf{0.946}   & \textbf{0.727}     & \textbf{0.960}     \\ \hline
\end{tabular}
}
\end{table}
\end{center}

\section{Experimental Results}
\label{sec:exp}

\begin{table*}[t]
\caption{Performance of the TG task. The results of \ours are reported separately for pre-training on \textsc{CodeSearchNet} or \textsc{Code2Seq}.}
\label{tab:tg}
\begin{center}
\begin{tabular}{c|cccc|cccc}
\hline
\multirow{2}{*}{Model}                                            & \multicolumn{4}{c|}{\textsc{JavaRef}} & \multicolumn{4}{c}{\textsc{TL-CodeSum}} \\ \cline{2-9} 
                  & Accuracy  & EM     & CER     & ED     & Accuracy  & EM     & CER      & ED      \\ \hline
\textsc{Naturalize}                                               & 0.035     & 0.029  & 336.356 & 10.793 & 0.067     & 0.059  & 242.757  & 10.762  \\ \hline
\begin{tabular}[c]{@{}c@{}}\ours\\ (\textsc{CodeSearchNet})\end{tabular} & 0.581     & 0.524  & 57.897  & 3.185  & 0.576     & 0.530  & 58.521   & 2.991   \\ \hline
\begin{tabular}[c]{@{}c@{}}\ours\\ (\textsc{Code2Seq})\end{tabular}    & 0.584     & 0.537  & 55.122  & 3.168  & 0.572     & 0.531  & 54.851   & 2.938   \\ \hline
\end{tabular}
\end{center}
\end{table*}

\begin{table*}[t]
\caption{Results of the ablation study on two datasets for the TG task. All methods are pre-trained on \textsc{CodeSearchNet}. Best results are shown in bold.}
\label{tab:ablation}
\begin{center}
\begin{tabular}{c|cccc|cccc}
\hline
\multirow{2}{*}{\textbf{Model}} & \multicolumn{4}{c|}{\textsc{JavaRef}}                              & \multicolumn{4}{c}{\textsc{TL-CodeSum}}                            \\ \cline{2-9} 
 & Accuracy       & EM             & CER             & ED             & Accuracy       & EM             & CER             & ED             \\ \hline
\oursCE           & 0.712          & 0.668          & 36.930          & 2.013          & 0.673          & 0.633          & 41.866          & 2.117          \\
\oursCECL                & 0.739          & 0.699          & 33.472          & 1.856          & 0.674          & 0.634          & 41.479          & 2.120          \\
\oursCEBOT                  & 0.740          & 0.697          & 31.817          & 1.804          & 0.692          & 0.653          & 39.836          & 2.015          \\
\oursBOTCL                & 0.040          & 0.024          & 110.750         & 7.829          & 0.230          & 0.198          & 104.027         & 5.610          \\ \hline
\ours                          & \textbf{0.750} & \textbf{0.711} & \textbf{31.164} & \textbf{1.775} & \textbf{0.700} & \textbf{0.662} & \textbf{37.834} & \textbf{1.975} \\ \hline
\end{tabular}
\end{center}
\end{table*}

In this section, we report and analyze the experimental results to answer the following research questions (RQ):
\begin{itemize} 
\item \textbf{RQ1:} Is \ours capable of predicting the correct number of tokens in the refactored variable name?

\item \textbf{RQ2:} Can \ours generate high-quality tokens in refactored variable names for rename refactoring?

\item \textbf{RQ3:} Does each component of \ours contribute to its performance?

\item \textbf{RQ4:} Do different settings affect the performance of \ours?
\end{itemize}

\subsection{Performance of the LP Task (RQ1)}

Tab.~\ref{tab:lp} shows the performance of the heuristic-based method and \ours on the LP task.
From Tab.~\ref{tab:lp}, we have several observations:
\begin{enumerate}

\item On both \textsc{JavaRef} and \textsc{TL-CodeSum}, \ours outperforms the heuristic-based method by 0.13-0.20 on Hit@1. 
Overall, \ours can correctly predict the number of tokens in the refactored variable name with a high probability (0.65-0.72).

\item The performance of \ours is around 0.94-0.96 on Hit@3 and it consistently exceeds the heuristic-based method by a large margin.
The high Hit@3 of \ours shows that it is very likely for \ours to predict the appropriate token number with the near-the-top rank.

\end{enumerate}
Based on the above observation, we can conclude that \ours is capable of predicting the correct number of tokens in the refactored variable name.

\subsection{Performance of the TG task (RQ2)}

\ours uses the top-$1$ predicted number in the LP task to guide itself to generate the corresponding quantity of tokens in the TG task.
Tab.~\ref{tab:tg} presents the performance of \ours and \textsc{Naturalize} on the TG task.

We can see that \ours significantly outperforms \textsc{Naturalize} on both \textsc{JavaRef} and \textsc{TL-CodeSum}.
There are two possible reasons:
\begin{enumerate}

\item Firstly, \textsc{Naturalize} only generates variable names that exist in the training data.
Due to the sparsity issue of variable names shown in Fig.~\ref{fig:var_fre}, it is very likely that the refactored variable name does not exist in the vocabulary of \textsc{Naturalize}. 

\item Besides, \textsc{Naturalize} offers variable-level rename refactoring and it does not capture the semantics of sub-tokens, which significantly affects its performance w.r.t. token-level metrics Accuracy, CER and ED. Differently, \ours is a token-level rename refactoring method and does not only generate existing variable names for rename refactoring.
Hence, it shows much better results on the TG task.

\end{enumerate}
In summary, we can conclude that \ours can generate high-quality tokens in refactored variable names for rename refactoring.

\subsection{Ablation Study (RQ3)}
\label{sec:as}

Since the token number predictor (Sec.~\ref{sec:pred_number}) used for the LP task is a simple single-layer feedforward neural network without additional components, we only conduct ablation study for the TG task.
To investigate whether each component in \ours takes effect for the TG task, we investigate the performance of the following variants of \ours:
\begin{itemize}
\item \textbf{\oursCE:} It only keeps $\mathcal{L}_{\text{cMLM}}$ in Eq.~\ref{eq:fine-tune-loss} during fine-tuning.

\item \textbf{\oursCECL:} It removes $\mathcal{L}_{\text{BoT}}$ in Eq.~\ref{eq:fine-tune-loss} during fine-tuning.

\item \textbf{\oursCEBOT:} It removes $\mathcal{L}_{\text{CL}}$ in Eq.~\ref{eq:fine-tune-loss} during fine-tuning.

\item \textbf{\oursBOTCL:} It removes $\mathcal{L}_{\text{CE}}$ in Eq.~\ref{eq:fine-tune-loss} during fine-tuning.

\end{itemize}

Tab.~\ref{tab:ablation} reports the results of the ablation study. All methods are pre-trained on \textsc{CodeSearchNet}.
All methods generate tokens according to the ground-truth token number in the variable name.
From the results, we have the following observations:
\begin{itemize}

\item We can see that \oursBOTCL shows the worst performance among all variations of \ours, meaning that completely removing the cross-entropy loss will significantly downgrade the performance on the TG task.
Hence, we keep $\mathcal{L}_{\text{cMLM}}$ in Eq.~\ref{eq:fine-tune-loss} to balance the strict order consistency brought by $\mathcal{L}_{\text{cMLM}}$ and the token order flexibility brought by $\mathcal{L}_{\text{BoT}}$.

\item All four variations of \ours show worse performance than \ours, showing the effectiveness of each component in \ours. Moreover, \oursCE performs worse than \oursCECL and \oursCEBOT, which further demonstrates that both the BoT loss and contrastive learning play important roles in \ours.

\end{itemize}

\vspace{5pt}
\noindent\textbf{Case Study.}
We provide 6 cases of rename refactoring on \textsc{JavaRef} in Fig.~\ref{fig:casestudy} as case study to illustrate the superiority of using the BoT loss over solely using the standard cross-entropy loss $\mathcal{L}_{\text{cMLM}}$ in Eq.~\ref{eq:fine-tune-loss} on the rename refactoring task. 
We observe that, if results using BoT are of low quality, results using $\mathcal{L}_{\text{cMLM}}$ only are also of inferior quality. 
Hence, we selected six cases where \ours produces consistent results as ground truth, in three categories (more meaningful, abbreviated, and non-repetitive).
From Fig.~\ref{fig:casestudy}, we can see that \ours using the BoT loss can generate more accurate variable names:
\begin{itemize}
\item \ours can produce more meaningful tokens according to the context of the variable name.
For instance, in Example 2, \textit{count} generated by using the BoT loss is a more meaningful variable name than \textit{i} generated by using the cross-entropy loss.

\begin{figure}[t]
    \centering
    \includegraphics[width=0.95\columnwidth]{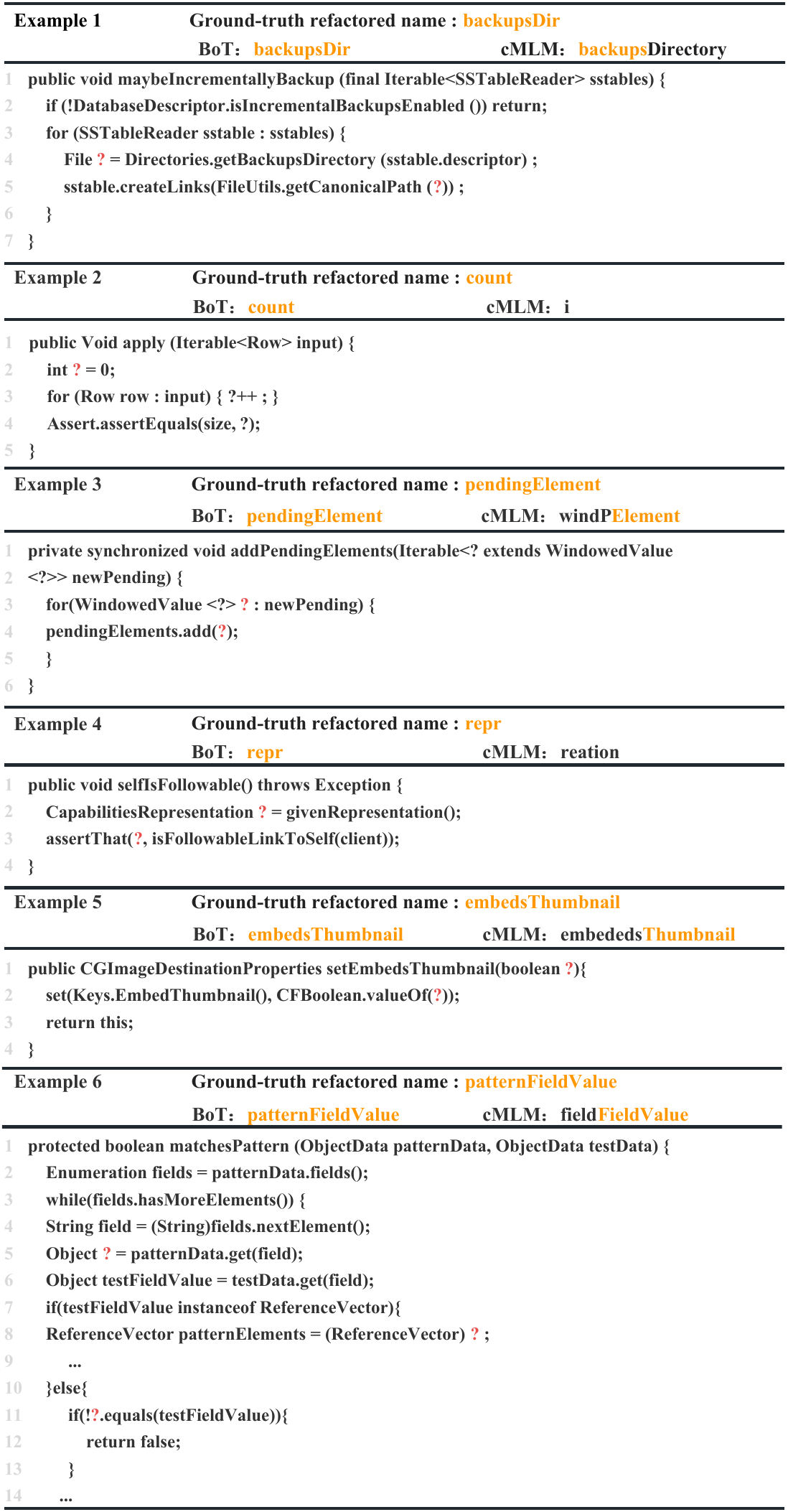}
    \caption{Case study for comparing using BoT loss and solely using the cross-entropy loss. Question marks indicate target variables of rename refactoring. Correct tokens are shown in yellow.}
    \label{fig:casestudy}
\end{figure}

\begin{figure}[t]
    \centering
    \subfloat{\includegraphics[width=0.46\columnwidth]{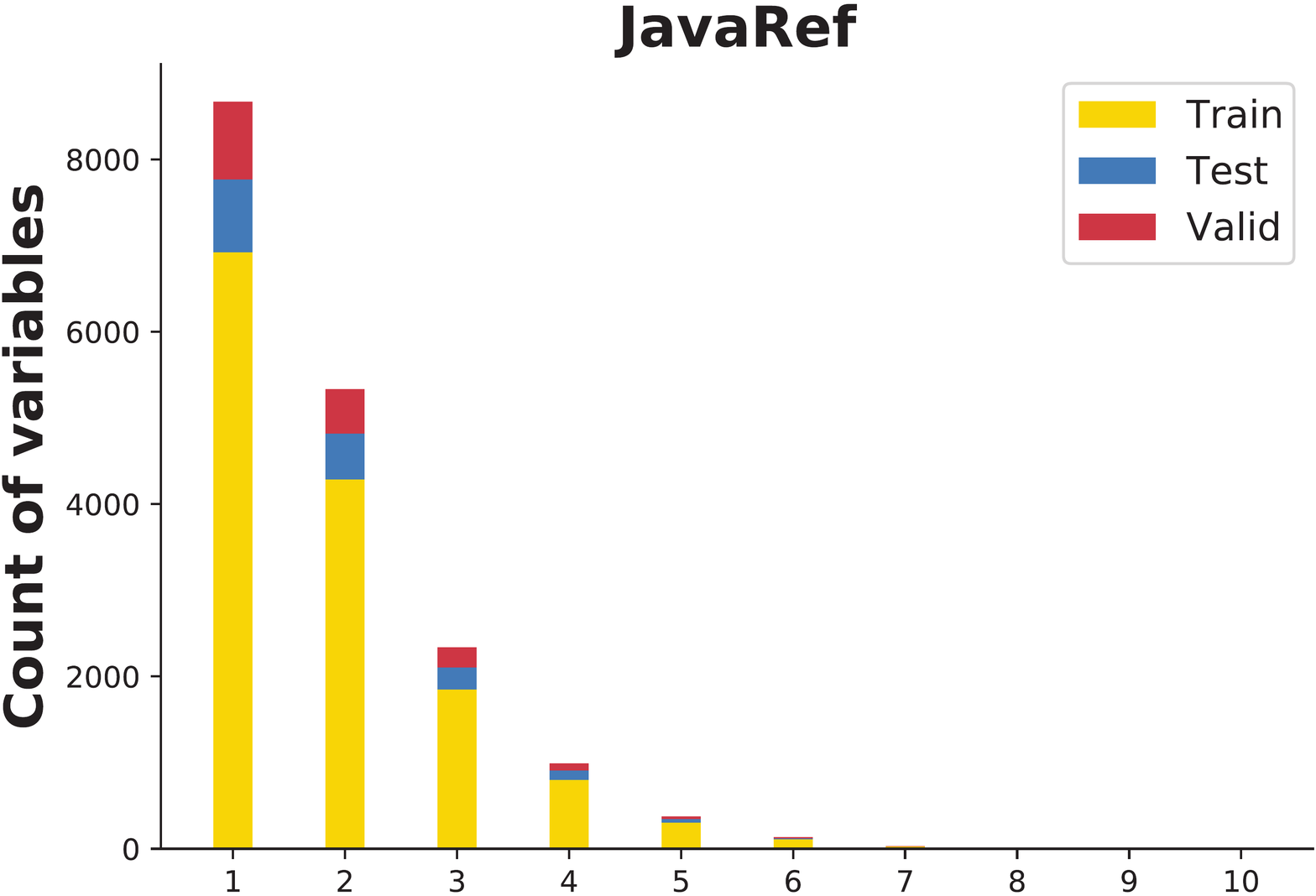}}
    \qquad
    \subfloat{\includegraphics[width=0.46\columnwidth]{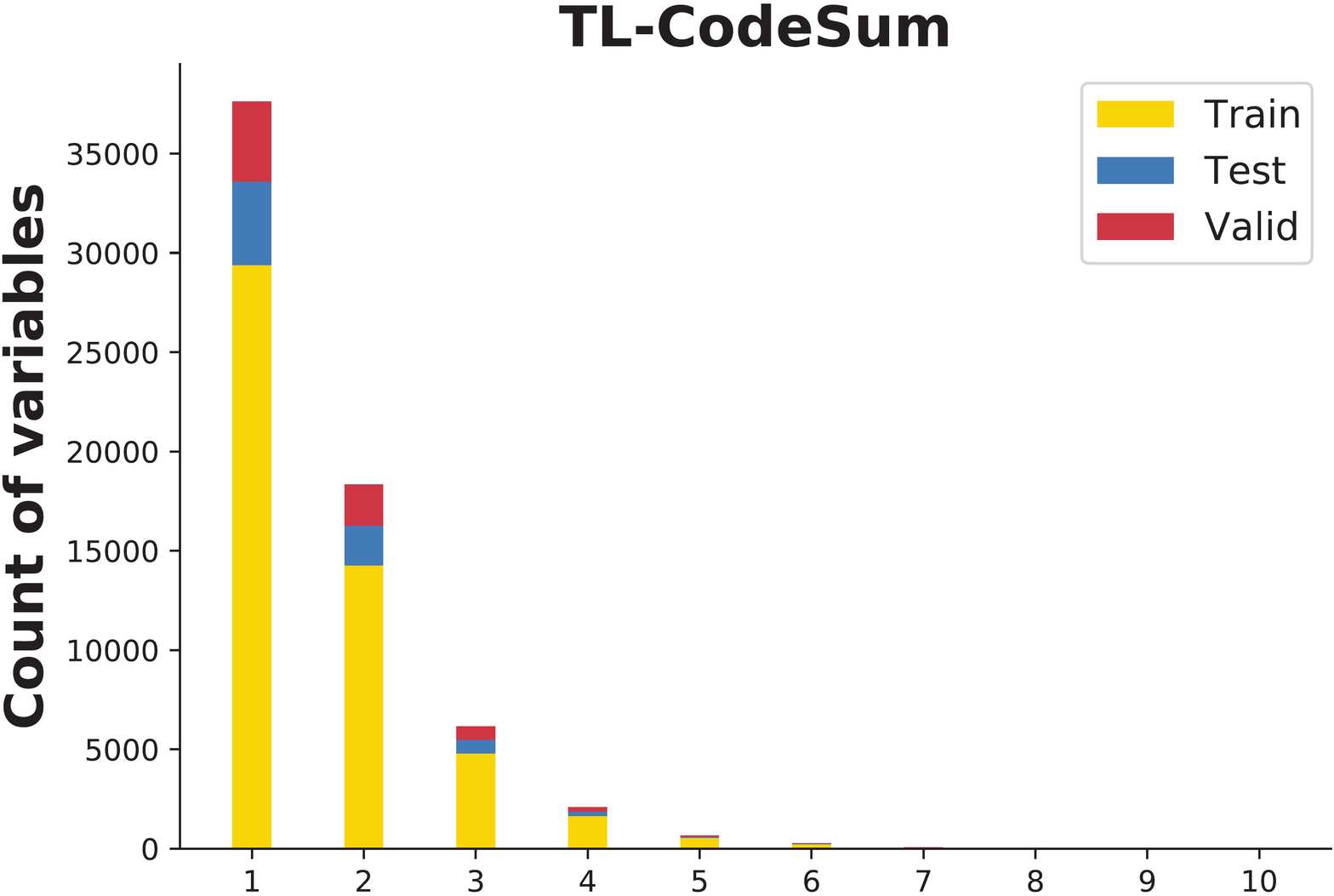}}
    \caption{Distributions of the number of tokens in variable names.}
    \label{fig:lengthdistribution}
\end{figure}

\begin{figure}[t]
    \centering
    \includegraphics[width=0.47\linewidth]{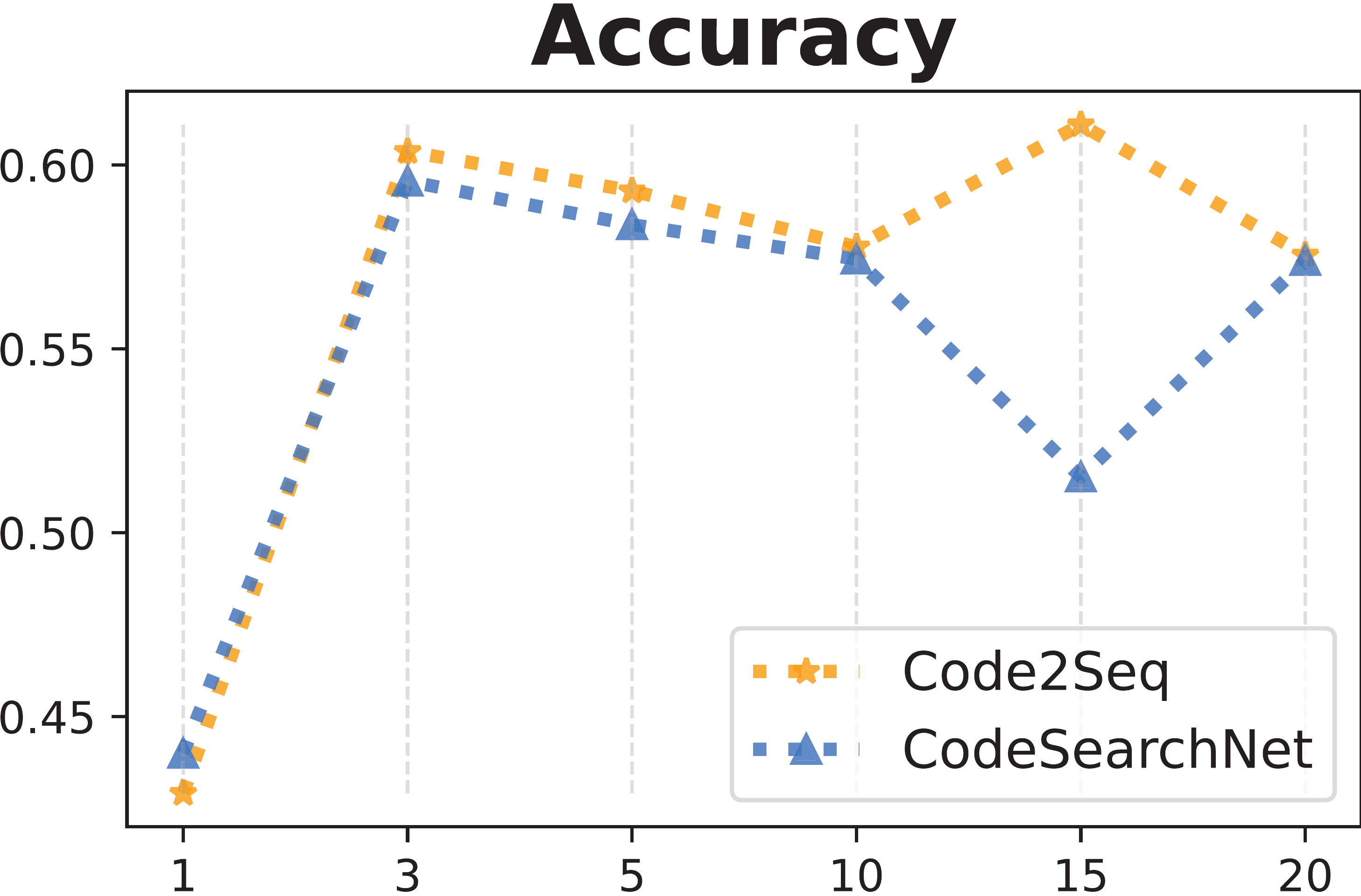}
    \hfill
    \centering
    \includegraphics[width=0.47\linewidth]{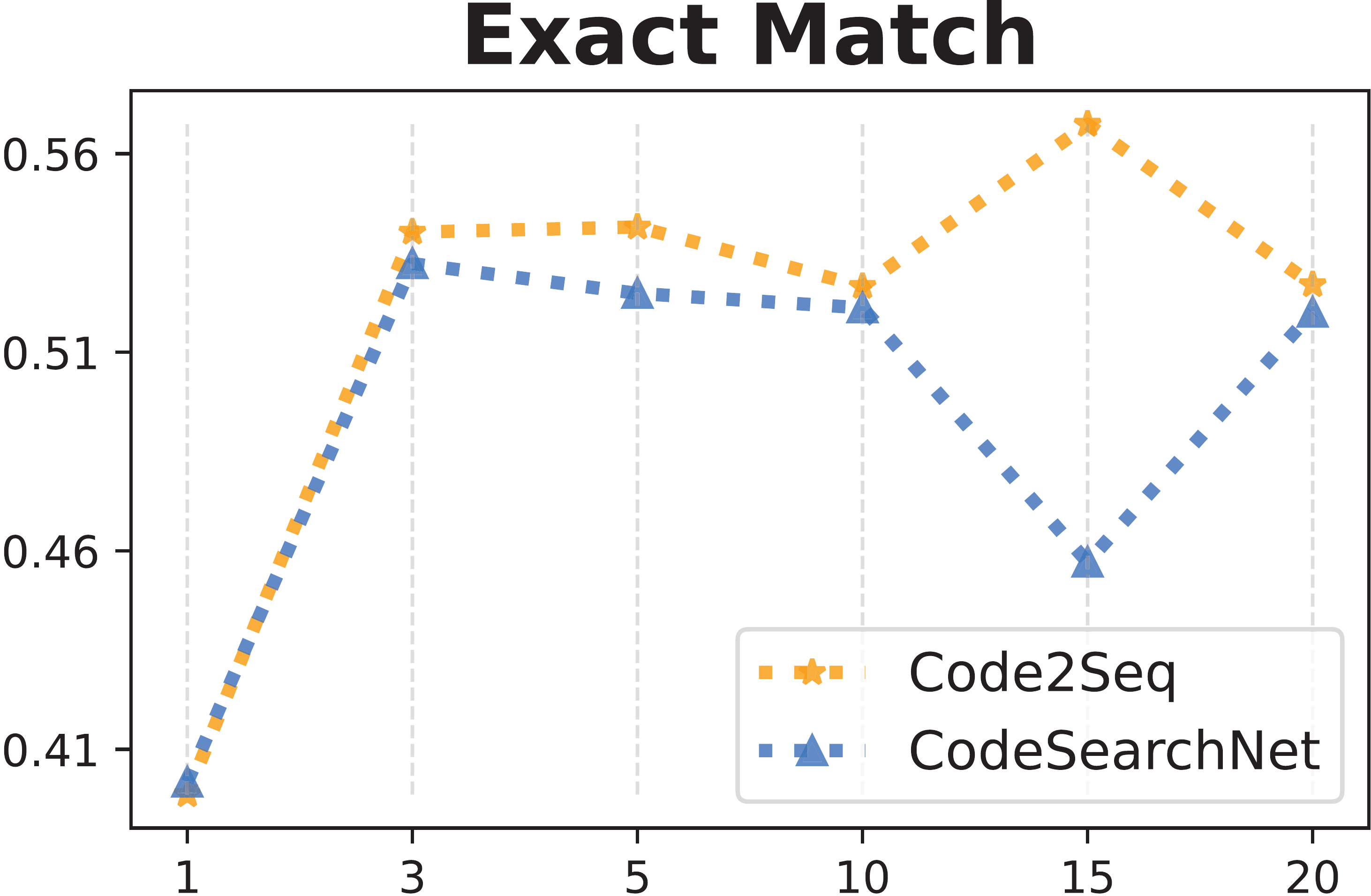}
    \caption{Effects of $l_{max}$ on the performance of \ours on \textsc{JavaRef}. Orange dotted line and blue dotted line indicate pre-trained \ours on \textsc{Code2Seq} and \textsc{CodeSearchNet}, respectively.}
    \label{fig:result_change_len}
\end{figure}

\item Moreover, using the BoT loss helps generate abbreviated variable names (e.g., \textit{Dir} for \textit{Directory} in Example 1 and \textit{repr} for \textit{Representation} in Example 4).
The reason is that the BoT loss cares the global probability of a token when generating tokens.
Assume \emph{backupsDir} is split into \emph{backups} and \emph{Dir}.
At step 1, the probability of \emph{backups} is highest followed by \emph{Dir}, while the probability of \emph{Directory} is quite low.
At step 2, the probability of \emph{Directory} is highest followed by \emph{Dir} and \emph{backups}.
Using the cross-entropy loss, the model is trained to generate the token at each step with the highest probability and hence it yields \emph{backupsDirectory}.
Differently, using the BoT loss, the model finds the global probability (i.e., consider probabilities for all steps) of \emph{Dir} and \emph{backups} are higher than \emph{Directory}.
As the predicted token number is 2, the model generates \emph{backupsDir} as the output.
This may be a desirable merit for the case where developers prefer abbreviations.

\item Using the BoT loss, \ours can avoid generating duplicate tokens. For example, solely using cross-entropy loss, \ours produces \textit{fieldFieldValue} in Example 6. As a comparison, using the BoT loss, the generated variable name is \textit{patternFieldValue}. The reason is that no duplicated tokens are recorded in the bag-of-tokens representation, and \ours generates $g_s^{(m)}$ unique tokens from the token vocabulary, as illustrated in Sec.~\ref{sec:predict}.

\end{itemize}

\begin{center}
\begin{table}[t]
  \begin{minipage}[b]{0.5\linewidth}
  \centering
  \caption{Impact of $\lambda$}
	\label{tab:lambda}
	\scalebox{0.9}{
        \begin{tabular}{@{}cccc@{}}
        \toprule
        $\lambda_{CL}$ & $\lambda_{cMLM}$ & $\lambda_{BoT}$ & Accuracy       \\ \midrule
        1              & 1                & 0               & 0.738          \\
        1              & 0                & 1               & 0.040          \\
        0.5            & 0.2              & 0.1             & 0.731          \\
        0.5            & 1                & 0.1             & 0.740          \\
        \textbf{1}     & \textbf{1}       & \textbf{0.1}    & \textbf{0.750} \\
        1              & 0.5              & 0.1             & 0.745          \\ \bottomrule
        \end{tabular}
	}
  \end{minipage}%
  \hfill
  \begin{minipage}[b]{0.5\linewidth}
  \centering
  \caption{Impact of $\tau$}
	\label{tab:tau}
	\scalebox{0.9}{
    \begin{tabular}{@{}cc@{}}
	\toprule
	$\tau$        & Accuracy      \\ \midrule
	0.02          & 0.741          \\
	\textbf{0.05}          & \textbf{0.750}          \\
	0.10          & 0.743          \\
	0.15          & 0.737          \\
	0.20          & 0.741          \\
	0.25          & 0.736          \\
	 
	\bottomrule
	\end{tabular}
	}
  \end{minipage}
\end{table}
\end{center}

\subsection{Effects of Different Settings (RQ4)}
\label{sec:effects}

\vspace{5pt}
\noindent\textbf{Impact of $l_{max}$.}
As explained in Sec.~\ref{sec:pred_number}, we formulate the prediction of the number of tokens in the refactored variable name as a multi-label classification task and the number of the labels is decided by the pre-defined maximum token number. 
In Fig.~\ref{fig:lengthdistribution}, we provide the distributions of the number of tokens in variable names in \textsc{JavaRef} and \textsc{TL-CodeSum}.
From Fig.~\ref{fig:lengthdistribution}, we can observe that more than 40\% of variable names contain only one token, more than 70\% of variable names contain no more than 2 tokens, and all variable names have no more than 10 tokens. 
To investigate the impact of using different $l_{max}$ on the TG task, we test \ours with different $l_{max}$ in $\{$1, 3, 5, 10, 15, 20$\}$ and report the results on \textsc{JavaRef} in Fig.~\ref{fig:result_change_len}.
From the results, we can see that the change trends of performance w.r.t. accuracy and exact match are mostly consistent on \textsc{JavaRef}.
Pre-trained \ours on \textsc{CodeSearchNet} shows the best result when $l_{max}=3$ or $l_{max}=5$.
Although pre-trained \ours on \textsc{Code2Seq} shows best result when $l_{max}=15$, most variable names do not contain many tokens as demonstrated in Fig.~\ref{fig:lengthdistribution}.
Therefore, our default setting $l_{max}=5$ is an appropriate choice in practice.

\vspace{5pt}
\noindent\textbf{Impact of $\lambda$.} Tab.~\ref{tab:lambda} reports the results using different $\lambda_{\text{CL}}$, $\lambda_{\text{cMLM}}$ and $\lambda_{\text{BoT}}$ in Eq.~\ref{eq:fine-tune-loss} on \textsc{JavaRef} for the TG task and \ours is pre-trained on \textsc{CodeSearchNet}.
The performance of the default setting is shown in bold.
We can see that, completely removing the cross-entropy loss ($\lambda_\text{cMLM}=0$) significantly affects \ours on the TG task, as already shown in the ablation study (Sec.~\ref{sec:as}). 
The result also provides proof for our previous idea of retaining $\mathcal{L}_\text{cMLM}$ while introducing $\mathcal{L}_{\text{BoT}}$ in Sec.~\ref{sec:generate_token}.
Moreover, from Tab.~\ref{tab:lambda}, we can conclude that \ours consistently shows \textasciitilde0.7 for accuracy on the TG task and our default setting yield the best result.

\vspace{5pt}
\noindent\textbf{Impact of $\tau$.} Tab.~\ref{tab:tau} provides the results using different $\tau$ in Eq.~\ref{eq:contrast_loss} on \textsc{JavaRef} for the TG task and \ours is pre-trained on \textsc{CodeSearchNet}. The performance of the default setting is shown in bold.
We can observe that the accuracy of \ours is not sensitive to the choice of $\tau$. Hence, it is easy to tune $\tau$.


\section{Threats to Validity}
\label{sec:threats}

Three threats may affect the validity of our study:

\vspace{5pt}
\noindent\textbf{Oracle Bias.} 
We construct \textsc{JavaRef} using \textsc{RefactoringMiner}, which may fail to identify some refactorings.
Moreover, using the refactoring detection tool, oracles may be biased towards refactoring instances that are easy for the tool to discover. 
It means, if multiple refactoring operations are performed on a code snippet at the same time, some refactoring types may be neglected as the tool may not detect all the instances. 
To mitigate oracle bias, we ask students with professional background to manually check the detected data and correct errors.

\vspace{5pt}
\noindent\textbf{False Oracles.} We adapt an existing dataset \textsc{TL-CodeSum} and use it for rename refactoring. We assume existing variable names in \textsc{TL-CodeSum} are meaningful and do not require rename refactoring. 
This may be true for most existing variable names, but some variable names may not be meaningful and require refactoring.
Thus, \textsc{TL-CodeSum} may contain false oracles. 
However, as the primary challenge for our task is the lack of data, we have to adapt \textsc{TL-CodeSum} in addition to \textsc{JavaRef} and hold such an assumption. 
As we are keeping enriching \textsc{JavaRef}, we expect \textsc{JavaRef} will be sufficient for the future study.

\vspace{5pt}
\noindent\textbf{Metrics.} 
To our best knowledge, there is no universally acknowledged metric for evaluating rename refactoring.
We choose several prevalent metrics from information retrieval and natural language processing domains for evaluation.
But there exists a threat that using one metric may not fairly evaluate the generated variable names.
Hence, we suggest considering all the metrics together to get a full picture of the performance of rename refactoring.


\section{Conclusion}
\label{sec:con}
Automatic rename refactoring reduces the intellectual burden of developers and improves the readability of programs.
However, few existing works express concern on how to suggest a new name for rename refactoring on variable names.
In this paper, we study automatic rename refactoring and particularly focus on refactoring variable names that are more difficult to refactor compared to other identifiers.
Based on our observations on rename refactoring, we propose \ours, a two-stage pre-trained framework tailored for rename refactoring.
Experimental results demonstrate the effectiveness of \ours.
In the future, we will explore how we can better determine the order of the generated tokens for constructing the refactored name.
Moreover, we will continue to construct and improve our refactoring dataset \textsc{JavaRef} to alleviate oracle bias and reduce false oracles.

To enhance reproducibility and replicability, we provide our implementation and data at \url{https://github.com/KDEGroup/RefBERT}.

\section*{Acknowledgments}
This work was partially supported by National Key R\&D Program of China (No. 2022ZD0118201), National Natural Science Foundation of China (No. 62002303, 42171456),  Natural Science Foundation of Fujian Province of China (No. 2020J05001) and CCF-Tencent Open Fund.

\balance
\bibliographystyle{ACM-Reference-Format}
\bibliography{ref}

\end{document}